\documentclass[prb,twocolumn,superscriptaddress,showpacs,amsmath,amssymb,floatfix]{revtex4}
\usepackage[T1]{fontenc}
\usepackage{verbatim}
\usepackage{amssymb}
\usepackage{graphicx}

\begin{document}

\title{Dynamical Anderson transition in one-dimensional periodically kicked incommensurate lattices}

\author{Pinquan Qin}
\affiliation{Beijing National Laboratory for Condensed Matter
Physics, Institute of Physics, Chinese Academy of Sciences,
Beijing 100190, China}

\author{Chuanhao Yin}
\affiliation{Beijing National Laboratory for Condensed Matter
Physics, Institute of Physics, Chinese Academy of Sciences,
Beijing 100190, China}

\author{Shu Chen}
\email{schen@aphy.iphy.ac.cn}
%\thanks{Corresponding author, schen@aphy.iphy.ac.cn}
\affiliation{Beijing National
Laboratory for Condensed Matter Physics, Institute of Physics,
Chinese Academy of Sciences, Beijing 100190, China}
\affiliation{Collaborative Innovation Center of Quantum Matter, Beijing, China}

%\date{}

\begin{abstract}
We study the dynamical localization transition in a one-dimensional periodically kicked incommensurate lattice, which is created by perturbing a primary optical lattice periodically with a pulsed weaker
incommensurate lattice. The diffusion of wave packets in the pulsed optical lattice exhibits either extended or localized behaviors, which can be well characterized by the mean square displacement and the spatial correlation function. We show that the dynamical localization transition is relevant to both the strength of incommensurate potential and the kicked period, and the transition point can be revealed by the information entropy of eigenfunctions of the Floquet propagator.
%The dynamical localization transition can be revealed through
%optical lattice.
%can be experimentally studied through the diffusion of wave packets in a one-dimensional incommensurate
%optical lattice.
\end{abstract}

\pacs{03.75.Lm, 72.15.Rn, 05.30.Rt}
%72.15.Rn %Localization effects
%03.65.Vf Phases: geometric; dynamic or topological
%71.10.Pm: Fermions in reduced dimensions
%03.75.Lm Tunneling, Josephson effect, Bose-Einstein condensates in periodic potentials, solitons, vortices, and topological excitations
%05.30.Rt, Quantum phase transitions
\maketitle
%{\it Introduction.-}
\section{Introduction}
As a fundamental phenomenon of quantum systems in the presence of disorder, Anderson localization has been found in a broad range of physical systems beyond the scope of traditional condensed matter physics \cite{Anderson,RMP-AT,AL50}, including light waves in photonic lattices and atomic matter waves in a one-dimensional (1D) disordered or quasi-periodic potential \cite{Billy,Roati,Lahini,Bermudez}. Particularly, for a Bose-Einstein condensate (BEC) trapped in a 1D quasi-periodic potential, it has been demonstrated that a transition from an extended state to an exponentially localized state exists with the change of the disorder strength \cite{Roati}. Although  most of studies on the Anderson localization focused on static disordered systems, the dynamic localization problem, which was originally put forward in the study of periodically kicked quantum rotors \cite{dl,Fishman,Casati90}, has also attracted much attention recently due to experimental realizations of the quantum kicked rotor in trapped cold atom systems interacting with pulsed standing wave of light \cite{Raizen} and the observation of Anderson localization in the kicked system \cite{Chabe,Ringot}.

As no external random element is introduced, the dynamic localization in the kicked rotor can be viewed as an analog of 1D Anderson localization in momentum space by mapping the system onto a quasi-random 1D Anderson model \cite{Fishman}. The effective randomness in the kicked rotor is rooted in mechanisms of incommensurability induced by the periodic driving, and consequently the localization for the kicked rotor occurs in momentum space, instead of real space as in the usual Anderson model. An interesting issue arose here is to study the interplay of periodic driving and disorder, which is not yet addressed in the previous study of static disorder systems and kicked rotor systems. To this end, we study the dynamic localization in a 1D optical lattice perturbed by an additional pulsed incommensurate lattice. Different from previous works \cite{Roati}, the disorder induced by the applied incommensurate potential is periodically added, and the system can be described by a periodically kicked Aubry-Andr\'{e} (AA) model. For the static AA model \cite{AA,Harper}, its eigenstates are either extended or localized
%with the transition point determined by the self-duality point
and a localization transition occurs by increasing the strength of incommensurate potential \cite{AA,koh1983,Ingolda}, which has been experimentally verified in a bichromatic optical lattice by observing the expansion dynamics of a trapped noninteracting BEC \cite{Roati}. While 1D static incommensurate optical lattices have been well studied \cite{Roscilde,Cai,Giamarchi,Deissler}, less attention has been paid on the pulsed incommensurate optical lattices.
In this work, we study the dynamical localization transition in the periodically kicked incommensurate lattice and find the dynamics is not solely determined by the strength of incommensurate potential, but also relevant to the driven frequency of the kicked potential. The tunability of the incommensurate
optical lattices \cite{Roati,Gadway} makes it feasible to experimentally
study the dynamical localization transition through the diffusion of wave packets in the pulsed 1D incommensurate
optical lattice.

%The experimental realization of the kicked rotor
%with laser-cooled atoms interacting with a pulsed standing
%wave allowed the first experimental observation of
%Anderson localization in 1D with atomic matter waves.
%

%{\it Model with periodically driven incommensurate potential.-}
\section{Model with periodically driven incommensurate potential}
We consider the model with periodically driven incommensurate
potentials described by the following Hamiltonian:
\begin{eqnarray}
H = \sum_{i} [ (-J \hat{c}_{i}^{\dag } \hat{c}_{i+1} + H.c.) + \sum_n \delta(t- nT) V_i \hat{n}_{i} ],
\label{Ham}
\end{eqnarray}
where  $\hat{n}_{i}=\hat{c}^\dagger_i \hat{c}_i$ is the particle
number operator and $\hat{c}^\dagger_i$ ($\hat{c}_i$) the creation
(annihilation) operator. Here $J$ is the nearest-neighbor
hopping amplitude and the incommensurate potential
\[
V_{i}=\lambda \cos(2\pi i \alpha)
\]
varies at each lattice site with $\alpha$ being an irrational
number and $\lambda$ the strength of the incommensurate potential. In contrast to the AA model \cite{AA} described by
$H = \sum_{i} [ (-J \hat{c}_{i}^{\dag } \hat{c}_{i+1} + H.c.) +  \lambda_{AA} \cos(2\pi i \alpha) \hat{n}_{i} ]$,
the on-site incommensurate potential in Eq.(\ref{Ham}) is periodically added with a pulsed period $T$. Because of this resemblance, we will refer to systems described by Eq.(\ref{Ham}) as the periodically kicked AA model.

Experimentally, the AA model can be realized by superposing two optical lattices with incommensurate frequency \cite{mod2009}. Similarly, the periodically kicked AA model may be realized by superimposing
two optical lattices of the form
\begin{equation}
V(x)  =  V_{1}(x)+V_{2}(x)\sum_{n}\delta(t-nT)\label{eq:vx}
\end{equation}
with $V_{1}(x)=s_{1} E_{R_1} \sin^{2}(k_{1}x)$ and $V_{2}(x)=s_{2} E_{R_2} T \sin^{2}(k_{2}x)$,
where $k_{i}=2\pi/\lambda_{i}$
are the lattice wave-numbers and $s_{i}$ are the heights of the two lattices in units of their recoil energies $E_{R_i}=h^2/(2m\lambda_i)^2$.
%and $\delta$ is an arbitrary phase.
The potential $V_{1}(x)$ is used to create a primary lattice, that is weakly perturbed by adding $V_{2}(x)$ periodically when
time equals multiples of the kicked period.
%\emph{\textbf{In the tight-binding limit, the strength of the incommensurate potential $\lambda$ derived from
%$V(x)$ has different dimension from $\lambda_{AA}$ in the AA model.}}
%to form the kicked driven potential $V(x)$.
%In the tight-binding limit, such a system can be mapped into the
%periodically kicked AA model with $\alpha=k_{2}/k_{1}$,
%$J=\left(4/\sqrt{\pi}\right)s_{1}^{0.75}\exp(-2\sqrt{s_{1}})$ and
%$\lambda=(s_{2} \alpha^2/2) \exp(-\alpha^{2}/\sqrt{s_{1}})$.
%We shall take $\alpha=(\sqrt{5}-1)/2$ and set $J=1$ as the unit of energy.
We note that the periodically kicked AA model is also related to the kicked Harper model \cite{Leboeuf} and
thus our scheme in terms of periodically added incommensurate optical lattices also provides a possible physical realization
of kicked Harper model.

\section{Dynamic evolution and dynamic Anderson transition}
\label{sec:evolu}

%{\it Dynamic evolution and dynamic Anderson transition.-}
The dynamical evolution of the periodically kicked system is determined by the Floquet unitary propagator \cite{gri1998} over one period,
which can be written as $U(T,0)=\exp\left(-iH_{0}T\right)\exp\left(-i\sum_{j}^{L}V_{j}\hat{c}_{j}^{\dag}\hat{c}_{j}\right)$,
where $H_{0}=-\sum_{j} \left(\hat{c}_{j}^{\dag} \hat{c}_{j+1}+H.c.\right)$ and $L$ is the lattice size. For convenience, we have set $\hbar=1$ and $J=1$ as the unit of the energy.
%(Note, we have taken the periodic boundary condition of the lattice in our research, namely, $c_{M+1}=c_{1}$).
Given an initial state $\left|\psi(0)\right\rangle = \sum_{i=1}^{L} C_{i} \left|i \right\rangle $ at $t=0$, the evolution state after one kicked period is given by $\left|\psi(T)\right\rangle=U(T,0)\left|\psi(0)\right\rangle$, where $\left|i\right\rangle = \hat{c}^\dagger_i  \left| 0 \right\rangle $ represents the state with a particle located in the $i$-th site. To get the distribution function of the evolution state, we need calculate
the matrix element of the Floquet propagator $\left\langle i\left|U(T,0)\right|j\right\rangle $.
%(The single particle in the lattice is considered in our research).
%However, it's hard to compute $\exp\left(-iH_{0}T\right)\left|m\right\rangle $.
Representing $\left|\phi_{\mu}\right\rangle =\sum_{i}C_{i}^{\mu}\left|i\right\rangle $ as the $\mu$th eigenvector
of $H_{0}$ with the single particle eigenenergy $E^0_{\mu}$, i.e., $H_{0}\left|\phi_{\mu}\right\rangle =E^0_{\mu}\left|\phi_{\mu}\right\rangle $,
%The coefficient $C_{m}^{\mu}$ can be obtained through numeric or analytic calculation.
we can calculate the matrix element of the Floquet propagator via the expression of $\langle i|U(T,0)|j\rangle=\sum_{\mu}C_{i}^{\mu}C_{j}^{\mu*}e^{-i(E^0_{\mu}T+V_{j})}$. By applying the Floquet propagator repeatedly, the state after
$N$ periods can be written as $\left|\psi(NT)\right\rangle =\left[U(T)\right]^{N}\left|\psi(0)\right\rangle =\sum_{i=1}^{L}C_{i}(NT)\left|i\right\rangle$. Here $U(T) = U(T,0)$ and we have used the relation $U(T,0)=U(2T,T)= \cdots = U(nT,(n-1)T)$.
\begin{figure}
\includegraphics[width=1.0\columnwidth]{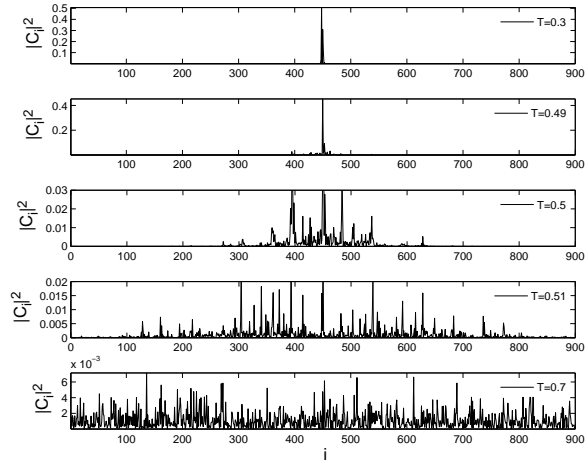}
\caption{The probability distribution of the state after $N=10^{4}$ periods in the periodically kicked AA model with $\lambda=1$. }
\label{cm-m}
\end{figure}
\begin{figure}
\includegraphics[width=1.0\columnwidth]{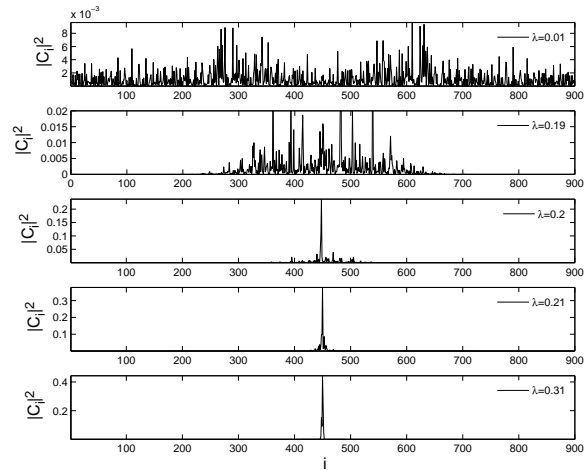}
\caption{The probability distribution of the state after $N=10^{4}$ periods in the periodically kicked AA model with $T=0.1$.}
\label{cm-m-1}
\end{figure}

For convenience, we take the initial state as $\left|\psi(0)\right\rangle =\left|L/2\right\rangle$, i.e., with the initial state located in the center of the lattice, and then study the expansion dynamics of the initial state in the pulsed incommensurate potential. To give a concrete example, in the following calculation we take $\alpha=(\sqrt{5}-1)/2$ and focus our study on the high-frequency regime with $1/T >1$.
It is known that the expansion dynamics on a static incommensurate lattice is only determined by the strength of incommensurate potentials, i.e., the evolution of the initial state exhibits quite different behaviors in the delocalization or localization regime \cite{Roati}. However, for the periodically kicked system, the expansion dynamics is determined by both the strength of incommensurate potentials and the driven frequency. To see it clearly, we first consider periodically kicked systems with the strength of the incommensurate potential fixed and variable driven frequencies.
Fixing the strength of the incommensurate potential at $\lambda=1$, we show distributions of expansion states after $N=10^4$ pulsed periods for systems with different driven periods $T=0.3, 0.49, 0.5, 0.51$ and $0.7$, respectively, in Fig.\ref{cm-m}. We can find that the final evolution state is still localized around the initial position when the driven period of the periodically kicked potential is smaller than a threshold, i.e., $T<0.5$. On the other hand, the final state expands to the whole lattice when the driven period is larger than a threshold.

Next we consider systems with the driven period of the
periodically kicked potential fixed and study the evolution dynamics for systems with different potential strengths. In Fig.\ref{cm-m-1}, we show distributions of expansion states after $N=10^4$ pulsed periods with the driven period fixed at $T=0.1$ for systems with different potential strengths. Our results clearly indicate that the evolution state is localized for $\lambda>0.2$, whereas it is extended when $\lambda<0.2$. Results shown in Fig.\ref{cm-m} and Fig.\ref{cm-m-1} indicate that the dynamic evolution of the periodically kicked systems is relevant to both the strength of incommensurate potentials and the driven frequency. The dynamic localization transition is determined by the ratio of $\lambda$ and $T$, i.e., the evolution state is either localized or extended  for $\lambda/T>2$ or $\lambda/T<2$.

%\section{mean square displacement}

\begin{figure}
\includegraphics[width=1.0\columnwidth]{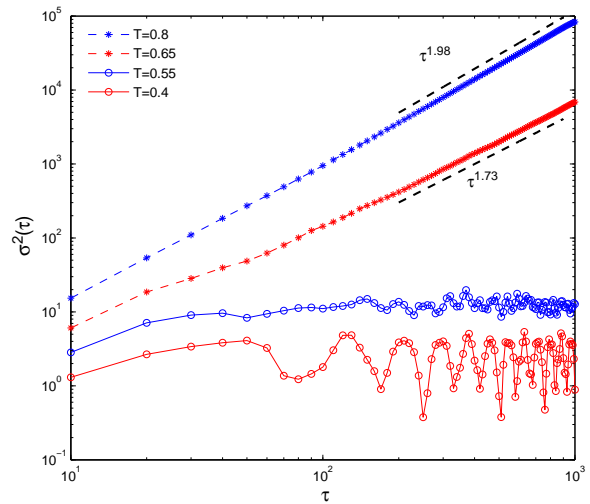}
\caption{Time dependence of $\sigma^{2}(\tau)$ in the periodically kicked AA model with $\lambda=1.2$ and $L=900$. The dash line represents a power-law fitting, which given $\sigma^{2}(\tau)\sim \tau^{1.98}$ with $T=0.8$, $\sigma^{2}(\tau)\sim \tau^{1.73}$ with $T=0.65$. }
\label{sigmat}
\end{figure}

To see how the wave packet spreads as a function of time, we calculate the mean
square displacement which is defined as \cite{zhang2012}
\[
\sigma^{2}(t)\equiv\sum_{i=1}^{L}(i-L/2)^{2}\left|C_{i}(t)\right|^{2}.
\]
%where $i$ represents the lattice site and $C_{i}(NT)$ depicts a normalized probability amplitude of the state after $N$ kicked period.
In general, during the expansion process, the mean square displacement increases as the power law of the time given by $\sigma^{2}(t)\sim t^{\gamma}$. The parameter $\gamma$ takes different values for the expansion in different lattices, for example, $\gamma=2$ in uniform lattices; $\gamma=0$ in disordered lattices.
%which also indicates a localized state.
While $\gamma=2$ and $\gamma=0$ correspond to ballistic diffusion and localization, respectively, the super-diffusion ($1<\gamma<2$) and sub-diffusion ($0<\gamma<1$) can occur in quasi-periodic lattices. In Ref.\cite{huf2001,zhang2012}, the quantum hyper-diffusion ($\gamma>2$) was also discovered.
For the kicked driven AA model, one can expect that the diffusion process is quite different for $\lambda/T>2$ or $\lambda/T<2$. To see it clearly, we calculate the mean square displacement as a function of time in units of the driven period with $\lambda$ fixed for different periods $T$.
In Fig.\ref{sigmat}, we show distributions of the mean square displacement $\sigma^{2}(\tau)$ with the strength of the periodically kicked potential fixed at $\lambda=1.2$ for systems with $L=900$ and different driven periods $T=0.4,0.55,0.65$ and $0.8$, respectively. For convenience, we have defined $\tau=t/T$. It is clear that the time-dependent mean square displacement displays different behaviors for $T>0.6$ or $T<0.6$. While the mean square displacement shows a power-law increase for $T=0.8$ and $T=0.65$, it oscillates around a given value after some expansion time and has zero power-law index for $T=0.55$ and $T=0.4$.
%Just as indicated by Anderson\cite{ander1958}, such state is a localized state.
As shown in Fig.\ref{sigmat},  the long-time power-law increase of  $\sigma^{2}(\tau)$ can be approximately described by $\sigma^{2}(\tau)\propto\tau^{1.73}$ for $T=0.65$ and $\sigma^{2}(\tau)\propto\tau^{1.98}$ for $T=0.8$, respectively. These power-law indexes indicate that the dynamical expansion is a super-diffusion process, which is in contrast to the localization process with zero power-law index for the expansion with $T=0.55$ and $0.44$. The property of the mean square displacement also indicates the occurrence of dynamical localization transition when $\lambda/T$ exceeds a threshold.
%\section{mean Green function}
\begin{figure}
\includegraphics[width=1.0\columnwidth]{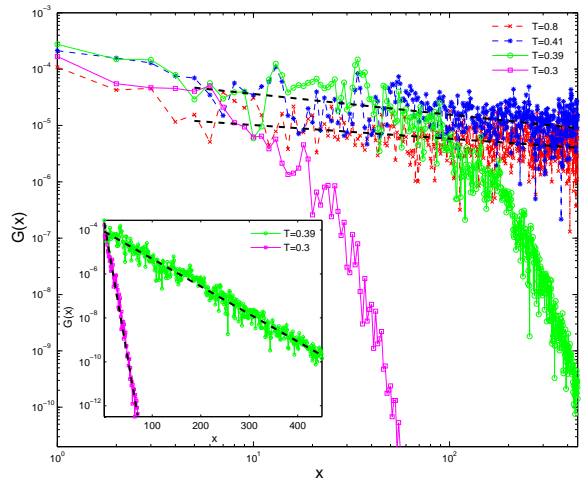}
\caption{The correlation function as a function of $x$ with $\lambda=0.8$, $L=2000$, $N=1.5\times10^{4}$ and various $T$.
The dash line indicates a power-law fit. In the inset, the dash line indicates an exponential-law
fit.}
\label{gxx}
\end{figure}

The dynamical localization can be also revealed by the correlation function
defined as, $G\left(x, t \right)\equiv L^{-1}\sum_{i}^{L}\left|\left\langle \psi(t)\left|c_{i}^{\dag}c_{i+x}\right|\psi(t)\right\rangle \right|
=L^{-1}\sum_{i}^{L}\left|C_{i}^{\ast}(t)C_{i+x}(t)\right|$, where $t=NT$.
Fixing the strength of the incommensurate potential at $\lambda=0.8$, we show distributions of the correlation functions after $N=1.5\times10^{4}$ pulsed periods for systems with different driven periods $T=0.3, 0.39, 0.41, 0.8$, respectively, in Fig.\ref{gxx}.
Our results indicate that the correlation function exhibits a power-law decay when the driven period of the periodically kicked potential is larger than a threshold, for examples, $G(x)\propto x^{-0.368}$ for $T=0.41$ and $G(x)\propto x^{-0.243}$ for $T=0.8$. On the other hand, the correlation function has an exponential-law decay when the driven period is smaller than the threshold, for examples, $ G(x)\propto e^{-0.0289 x}$ for $T=0.39$ and $G(x)\propto e^{-0.292 x}$ for $T=0.3$ as shown in the inset of Fig.\ref{gxx}.
The exponential-law decay of the spatial correlation function is the characteristic of the system in a dynamical localized state.

%\section{information entropy}

We have demonstrated that the extended or localized property of the dynamic evolution state can be well characterized by the mean
square displacement and the spatial correlation function of the evolution state. Moreover, we find that the eigenfunction of the Floquet unitary propagator can be also used to determine the transition point from the dynamical extended state to localized state, which is irrelevant to the choice of the initial state.
%\begin{figure}
%\includegraphics[width=1.0\columnwidth]{e-delta.eps}
%\caption{The eigenphase spectrum of the Floquet propagator in the kicked driven
%Aubry-Andr$\acute{e}$ model with $T=0.49$, $\lambda=1$.}
%\label{euspec}
%\end{figure}
Given that $\left|\psi_{\eta}\right\rangle $ is the eigenstate of the Floquet propagator $U(T)$ with the Floquet energy $E_{\eta}$, i.e., $U(T)\left|\psi_{\eta}\right\rangle =e^{-iE_{\eta}T}\left|\psi_{\eta}\right\rangle $,
in the basis of $|i\rangle$, we can represent $\left|\psi_{\eta}\right\rangle =\sum_{i=1}^{L}C_{i}(E_{\eta})|i\rangle$.
%where $C_{i}(E_{U})$ are the probability amplitude.
Then one can introduce the information entropy \cite{IE,Amico} defined as
\[
S^{inf}_{\eta} \equiv- \sum_{i=1}^{L}\left|C_{i}(E_{\eta})\right|^{2}\ln\left|C_{i}(E_{\eta})\right|^{2}.
\]
The information entropy takes it's minimum $S^{inf}_{\eta}=0$, whenever the state is localized in a single site, while it takes it's maximum $S^{inf}_{\eta}=\ln(L)$, when the state is completely extended with the wave function probability amplitudes given by $|C_{i}(E_{\eta})|=1/\sqrt{L}$.

\begin{figure}
\includegraphics[width=1.0\columnwidth]{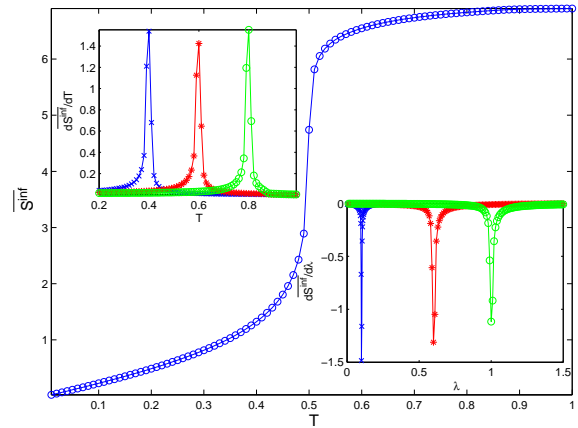}
\caption{The mean information entropy versus $T$ for the system with $\lambda=1$ and $L=1500$. The left up inset shows
the derivative of the mean information entropy versus $T$ with $\lambda=0.8$ (left
plot); $\lambda=1.2$ (middle plot); $\lambda=1.6$ (right plot). The right down
inset shows the derivative of the mean information entropy versus $\lambda$ with
$T=0.05$ (left plot); $T=0.3$ (middle plot); $T=0.5$ (right plot).}
\label{infentr}
\end{figure}

Fixing the strength of the incommensurate potential at $\lambda=1$, we show the mean information entropy of the Floquet unitary propagator versus the driven period $T$ in Fig.\ref{infentr}, where the mean information entropy is defined as $\overline{S^{inf}}\equiv L^{-1}\sum_{\eta=1}^{L} S^{inf}_{\eta}$.
%In the kicked driven Aubry-Andr$\acute{e}$ model, when we fixed the
%potential strength $V$, with the increase of the period of the potential $T$,
It shows that the mean information entropy increases from a tiny value to a finite large value with the increase of the pulsed period $T$, which indicates the wave function of the periodically kicked AA model undergoing a translation from localized state to extended state. In the up inset of Fig.\ref{infentr}, we show the derivative of the mean information entropy as a function of $T$ for systems with different potential strengths.
It turns out that the extremum of the derivative appears at $T=0.4$ for $\lambda=0.8$, $T=0.6$ for $\lambda=1.2$ and $T=0.8$ for $\lambda=1.6$. Similarly, in down inset of Fig.\ref{infentr}, the derivative of the mean information entropy as a function of $\lambda$ for systems with different pulsed periods is displayed, with extremum of the derivative located at $\lambda=0.1$  for $T=0.05$, $\lambda=0.6$ for $T=0.3$ and $\lambda=1.0$ for $T=0.5$. It is clear that the extremum of the derivative of mean information entropy appears at $\lambda/T=2$ for different systems, corresponding to the transition point from the dynamical localization to delocalization state.

\begin{figure}
\includegraphics[width=1.0\columnwidth]{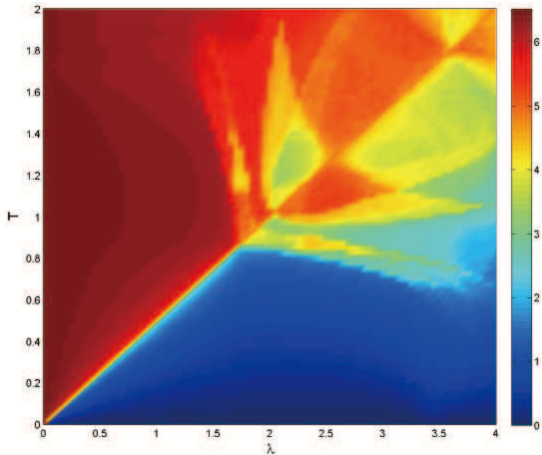}
\caption{(Color online) The mean information entropy versus both $\lambda$ and $T$ for the system with $L=900$.}
\label{inftv}
\end{figure}

Our numerical results indicate that, in the high-frequency regime, the dynamical localization transition point of the periodically kicked Aubry-Andr\'{e} model is located at $\lambda/T=2$. To understand this explicitly, we explore the effective Hamiltonian of the periodically kicked Aubry-Andr\'{e} model in the high-frequency regime. The effective Hamiltonian $H_{\mathrm{eff}}$ can be obtained from the Floquet unitary  propagator by the relation
 \[
  U(T) = \exp\left(-iH_{\mathrm{eff}}T\right) .
 \]
 As displayed in the appendix, the effective Hamiltonian is derived by using the Baker-Campbell-Hausdorff formula. In the high frequency and weak disorder limit with $1/T \gg 1$ and $\lambda \ll 1$, the effective Hamiltonian takes the form of
 \[
 H_{\mathrm{eff}} = -\sum_{i} \left(\hat{c}_{i}^{\dag} \hat{c}_{i+1}+H.c.\right)+\lambda/T \sum_i \cos(2\pi i \alpha) \hat{n}_i
 \]
 by omitting high-order terms which contain commutators. This effective Hamiltonian is just the static AA model with scaled potential strength $\lambda/T$, which indicates the localization transition point given by $\lambda/T=2$. However, in the low frequency regime, the high-order terms can't be omitted, and one can not expect that the dynamical localization transition can be described in the scheme of effective AA model. To see it clearly, we also calculate the mean information entropy in a larger parameter region, which is displayed in Fig.\ref{inftv} with the mean information entropy as a function of the strength of incommensurate potential and the kicked period. It's shown that there is a sharp change across the line of $\lambda/T=2$ in the high frequency region, similar to the specific case displayed in Fig.5. However, as shown in Fig.\ref{inftv}, the mean information entropy displays a more complicated distribution pattern when the system deviates the high frequency region. Consequently, the dynamic localization transition occurring at $\lambda/T=2$ breaks down when $1/T<1$.

In order to connect to the potential experimental realization, we refer our system to the parameter in Ref.\cite{Roati} in which $^{39}\mathrm{K}$ in a incommensurate lattices with height of the primary lattice $s_1=5$ in units of the recoil energy $E_{R1}=4785h\mathrm{Hz}$. Thus the hopping magnitude $J=323.6h\mathrm{Hz}$. Correspondingly, the period $T$ in unit of $\hbar/J$ is of the order of $10^{-4}\mathrm{s}$. Therefore the pulsed period  $T \sim 1 \mu \mathrm{s} ~-~ 10 \mu \mathrm{s}$ in typical kicked atom experiments \cite{Raizen,Ringot,Chabe,Gadway} is in the high-frequency regime discussed in the current work, which makes it possible to study the phenomena of the dynamical localization transition.
%In the real experiment, we also note the strength of the incommensurate potential of the periodically kicked AA model ($\lambda$) and the AA model %($\lambda_{AA}$) have different dimension. The dimension of $\lambda$ is $[J]\cdot[T]$.}

\section{summary}
%{\it Summary.-}
In summary, we have revealed the dynamical Anderson localization transition in a 1D periodically kicked incommensurate optical lattice by studying the diffusion of wave packets. The dynamical evolution of wave packets indicates that the dynamical state is either extended or localized, depending on both the strength of incommensurate potential and the kicked period.
We characterize the dynamical transition from various aspects by calculating the mean square displacement, the spatial correlation function and the information entropy of eigenfunctions of the Floquet propagator. These quantities all indicate the dynamical localization transition occurring at $\lambda/T=2$ in the high frequency regime, which can be also interpreted from the effective Hamiltonian of the system.
%In addition, the information entropy of eigenfunctions of the Floquet propagator also tell us there are no well-defined laws of the extended-localized %transition point in the low frequency strong disorder region.
Our observations and theoretical analysis should stimulate experimental studies of the phenomena of dynamical localization transition in the pulsed incommensurate optical lattices.

\section*{Acknowledgment}

This work has been supported by National Program for Basic Research of MOST, by NSF of China under Grants No. 11374354, No. 11174360, and No. 11121063, and by the Strategic Priority Research Program of the Chinese Academy of Sciences under Grant No. XDB07000000.

\appendix
\section{Derivation of effective Hamiltonian of periodically kicked Aubry-Andr\'{e} model}
\label{effh}
In Sec.\ref{sec:evolu}, the Floquet unitary propagator of the periodically kicked Aubry-Andr\'{e} model is written as a product of two exponential operators. From this propagator, we can derive the effective Hamiltonian of the kicked system, namely
\begin{eqnarray}
U(T) &=& \exp\left(-iH_{0}T\right)\exp\left(-i\lambda\hat{V}\right)\nonumber\\
&=& \exp\left(-iH_{\mathrm{eff}}T\right)
\label{heff}
\end{eqnarray}
where $\hat{V}=\sum_{j}^{L}\cos(2\pi i \alpha)\hat{c}_{j}^{\dag}\hat{c}_{j}$.
In the following analysis, we shall use the Baker-Campbell-Hausdorff (BCH) formula which determines $\hat{Z}$ such that $e^{\hat{A}}e^{\hat{B}}=e^{\hat{Z}}$ in the following way \cite{wil1967}
\begin{eqnarray}
e^{\hat{A}}e^{\hat{B}}&=&\exp\Bigg(\hat{A}+\hat{B}+\frac{1}{2}\left[\hat{A},\hat{B}
\right]+\frac{1}{12}\left[\hat{A},\left[\hat{A},\hat{B}\right]\right]\nonumber\\
&&+\frac{1}{12}\left[\left[\hat{A},\hat{B}\right],\hat{B}\right]+\cdots\Bigg)
\end{eqnarray}
Using this formula to Eq.(\ref{heff}), we derive the effective Hamiltonian as
\begin{eqnarray}
H_{\mathrm{eff}} &=& H_0+\frac{\lambda}{T}\hat{V}-i\frac{\lambda}{2}\left[H_0,\hat{V}\right]
-\frac{T\lambda}{12}\left[H_0,\left[H_0,\hat{V}\right]\right]\nonumber\\
&&-\frac{\lambda^2}{12} \left[\left[H_0,\hat{V}\right],\hat{V}\right]+\cdots.
\label{effha}
\end{eqnarray}
From the above expression, it is obvious that the effective Hamiltonian can be simplified as $H_{\mathrm{eff}} = H_0+\frac{\lambda}{T}\hat{V}$ in the limit of $1/T \gg1$ and $\lambda \ll 1$, which is just the AA model with $J=1$ and $\lambda_{AA} = \lambda/T$. However, one must consider the high-order terms when the system is not in the high-frequency regime.

\end{document}